# Executable Multi-Layered Software

*Lukas Radosky* lukas.radosky@fmph.uniba.sk ORCID: 0003-3909-3219

*Ivan Polasek* ivan.polasek@fmph.uniba.sk ORCID: 0000-0001-6004-701X

*Department of Applied Informatics, Faculty of Mathematics, Physics and Informatics, Comenius University. Mlynská dolina F1 Bratislava Slovakia 842 48*

## Abstract

This paper introduces a novel software visualisation and animation method, manifested in a prototype software tool - *AnimArch*. The introduced method is based on model fusion of static and dynamic models. The static model is represented by class diagram while the dynamic model is represented by source code written in high-level Object Action Language from xUML (executable UML). The class diagram defines architecture that is animated in response to real-time execution of the source code. Moreover, additional object diagram layer represents all object instances present in runtime. The *AnimArch* also features source code generation to Python, to bridge the gap from design to implementation. This paper provides detailed description of the modelling method and screenshots of the accompanying software tool.

## 1. Introduction

Software modelling visualisation is an important practice in overcoming the invisibility and complexity properties of software systems, as described by Brooks [1] The ability to perceive visually what does not have natural physical representation greatly enhances the ability to express, communicate, and understand its properties. Furthermore, graphical representations have been empirically proven to be more effective than textual descriptions [12, 13].

Since its introduction, the Unified Modelling Language (UML) has been the most widely adopted software visualisation method, in its nature static. However, there is no consensus in dynamic, i.e. animated software modelling techniques although they are subject to active research. Due to vast possibilities and purposes, perhaps there cannot ever be a unified approach to this problem, which is why we wish to contribute to the spectrum of ideas in this area.

To overcome the difficulties in explaining a software to newbies and newcomers, we propose a software modelling and visualisation method that simplifies transition from understanding software structure to understanding its implementation. By model fusion, the proposed modelling method enables to define a software model with class diagram and a set of source codes (corresponding to methods in the class diagram). We provide a software prototype, that provides execution of such model. Model execution has graphical output, visualising method invocation, object instantiation, assignment to object attributes and more. The prototype also enables source code generation, from the high-level platform-independent language defining the model source codes to platform-dependent language (only Python at the present). The method is intended not only to simplify introduction of new software development team members to the developed software, but we also believe it would help in teaching principles of object-oriented programming and design patterns.

The remainder of this paper is structured as follows. Section II provides overview of relevant research papers. Section III introduces the proposed modelling method and the supporting software tool. Section IV sums up the contribution of this paper and hints the future research directions of this work.

## 2. Related work

Much of the software modelling research revolves around UML diagrams and object-oriented programming. Savary-Leblanc and Le Pallec proposed a class diagram highlighting method [23]. Their prototype highlights classes associated to the currently selected class, enabling easier visual navigation, as indicated by their user evaluation. However, no dynamic animation is involved. Oberhauser proposed a UML visualisation and modelling method in VR [20]. Their prototype visualises several stacked layers with various UML diagrams, including use case diagram, component diagram and deployment diagram. Yigitbas et al. proposed a collaborative UML-based software modelling approach in VR [28,29]. UML class diagrams are composed of cuboids placed in planar space, with all six sides of cuboids containing identical description. Vesin et al. [26] proposed a collaborative modelling tool OctoUML that allows to create hand-drawn class and sequence diagrams. The crude shapes are converted into corresponding elements. Stanček et al. developed collaborative class diagram modelling environment in virtual reality [25], focusing on quality VR and user-to-user interaction over UML-related novelty.

As UML provides various diagrams, each offering a unique perspective on the modelled software, there has been some effort to merge these perspectives. Gregorovič et al. proposed software modelling approach where several UML diagram types, e.g. class, activity, and sequence diagrams stacked next to each other in layers [7], as shown in Fig 1a. This idea was further explored by Ferenc et al. who introduced collaborative modelling tool using similar approach [5], as shown in Fig. 1b. The idea of layering software engineering diagrams atop each other is however much older [6], although the research did not provide a supporting tool neither user evaluation at the time.

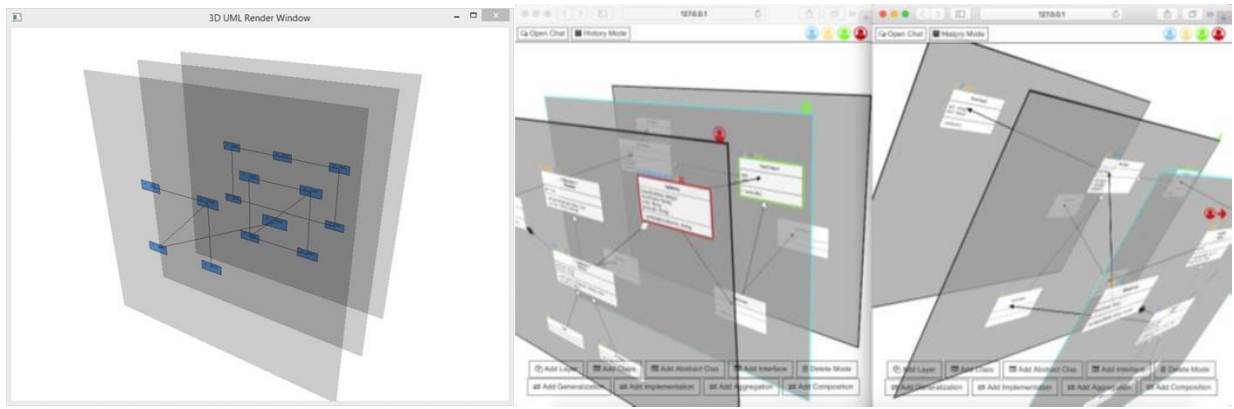

**Figure 1:** Multilayer modelling approaches proposed by (a) Gregorovič [7] and (b) Ferenc [5].

Often, the source code related to the modelled software is part of the visualisation and modelling approaches. Esteves and Mendes proposed dynamic class diagram, object diagram and source code visualisation method [4]. All views are displayed in IDE-like GUI. Yang et. al introduced both static (class diagram) and dynamic visualisation (object diagram) of Java source code [27]. Krause-Glau and Hasselbring introduced software visualization approach providing architectural overview of large Java projects within IDE that also enables source code editing [15]. Kučečka et al. [16] proposed a fusion of UML-based modelling and programming approach in VR. The approach is focused on 2D-based class diagram and related source code editing.

Other concepts of software engineering which are subject of research in regard to visualisation and animation are data structures and algorithms. Halim proposed a data structure and algorithm animation method [8]. Their focus was on well-known data structures such as linked list, binary heap, etc. Sparsha et al. have introduced procedural-programming-oriented approach by animating C source code and data structures [24]. Live animation of primitive variables and objects in Java environments was introduced by Hendrix et al [10]. Hayatpur et al. have introduced a JavaScript source code execution animation tool [9]. Pirker et al. introduced sorting algorithm visualisation and animation in virtual reality [21]. Kumar et al. introduced data structure visualization and animation tool [17].

Alternative approach to software visualisation is usage of metaphors. Sajaniemi et al. introduced workshop-based metaphor for early stages of teaching OOP [22]. Mills et al. introduced analogy for polymorphism, representing types with stacked shapes, where ability to fit shapes into each other indicates a type being a subtype of the other type [19]. Lian et al. proposed OOP learning method using a house metaphor, with doors representing access modifiers and floor for polymorphism [18]. House metaphor for software was also used by Hori et al. in virtual reality [11].

The general trend in user evaluation of novel software modelling approaches seems to be that these approaches improve model understanding and communication. As for problem solving, especially VR-based solutions seem to reduce effectiveness but improve user satisfaction. Most evaluations are performed on small software models, even thought it has been subject to criticism [14]. The argument for using small software models is that evaluation is meant to test the modelling tools, not the participants' intellectual abilities.

## 3. Our approach

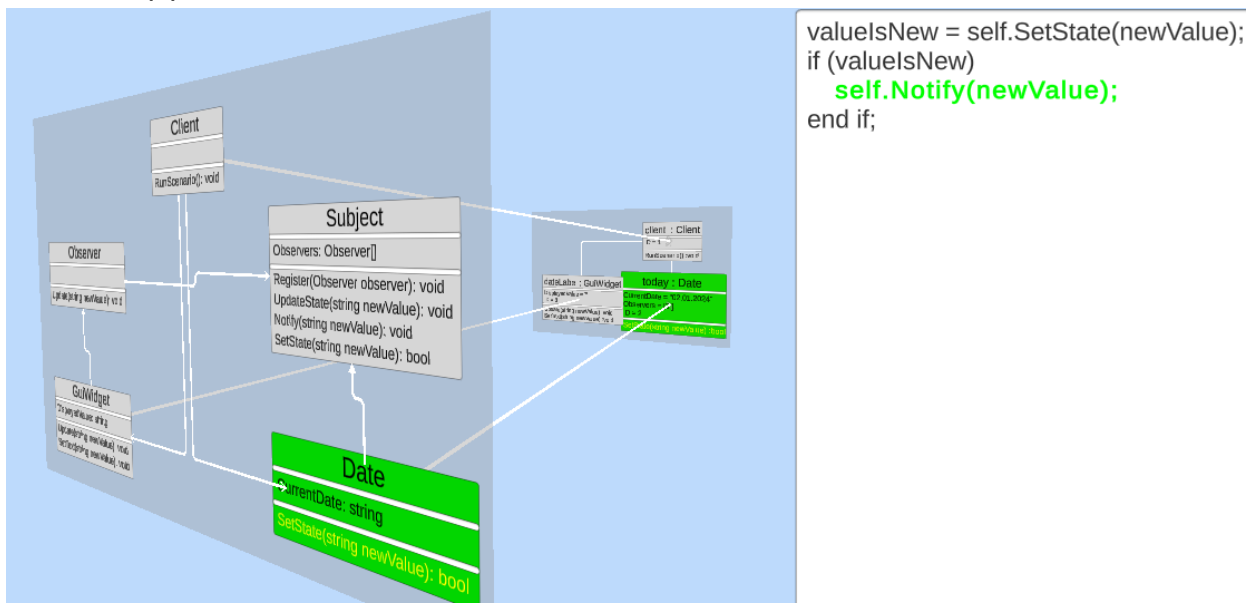

**Figure 2:** Overview of graphical user interface of our prototype.

We propose a software modelling method utilizing fusion of static and dynamic models. The static component here is the class diagram, the dynamic component is source-code-defined animation of the class diagram, and additional animated visualisations. The aim of this approach is to demonstrate source code execution by animating both the object-oriented aspect of the source code and the procedural aspect of individual methods. Visualisation consists of several UML diagrams stacked on each other and the related source code. The proposed method is supported by a software prototype, *AnimArch*. A screenshot of its graphical user interface is shown in Fig. 2.

*AnimArch* also features an option for collaborative class diagram editing. However, further effort is needed before the feature is presented, as collaboration should be enabled in all *AnimArch*'s features so that collaboration is possible during all phases of the proposed modelling method.

## 3.1. Defining a model

To define a model, both the static and the dynamic components need to be defined. Since the static component of the model is a class diagram, a class diagram needs to be drawn. *AnimArch* features its own class diagram editor. It offers simple click-based creation and modification of classes, relations, attributes, methods, and their parameters. The created class diagram can be saved to and loaded from files with in-house JSON schema. Alternatively, it is possible draw a class diagram in the well-known Enterprise Architect[1] software. The diagram then needs to be exported to an XMI file (XMI 2.1 format is required), and the file can be loaded via the *AnimArch*'s GUI.

The dynamic component of the model is source code written in customized Object Action Language[2] (OAL) syntax. OAL is used as an action language in executable UML[3] [2] (xtUML or xUML), therefore it is platform-independent. This is also the reason why we chose OAL over other languages - the platform independence enables us to generate platform-specific source code from our model. We only utilize a subset of OAL syntax but respect its semantics. Over time, we modified syntax of some commands, e.g. method invocation command, to enable users unfamiliar with OAL but familiar with more well-known languages to work with *AnimArch* effectively. One of the motivations here is the future evaluation - we will want to evaluate users' interaction with *AnimArch* rather than their ability to learn a new programming language. *AnimArch* features a simple text editor to create and edit source code of individual methods. The source code and be stored in JSON files with in-house schema.

There is no explicit connection between a file containing the class diagram and a file containing method source codes. It is therefore possible to load a class diagram file with various source code files, and vice versa. The idea behind this is not to force users to indicate various scenarios of animation explicitly in the class diagram, e.g. by creating a class *Scenarios* with methods *Scenario1*, *Scenario2*, etc., should there be several desired scenarios.

## 3.2. Executing a model

Model is executed and animated in *AnimArch*'s GUI. To execute and animation, it is necessary to select a starting method first. Any animation (execution) can start in any of the methods for which it defines a body, i.e. any method that has at least one command in it. Each method's source code is parsed via ANTLR[4]-generated parser into in-memory abstract syntax tree representation.

---

[1] https://sparxsystems.com/
[2] http://www.ooatool.com/docs/OAL08.pdf
[3] https://xtuml.org
[4] https://www.antlr.org/

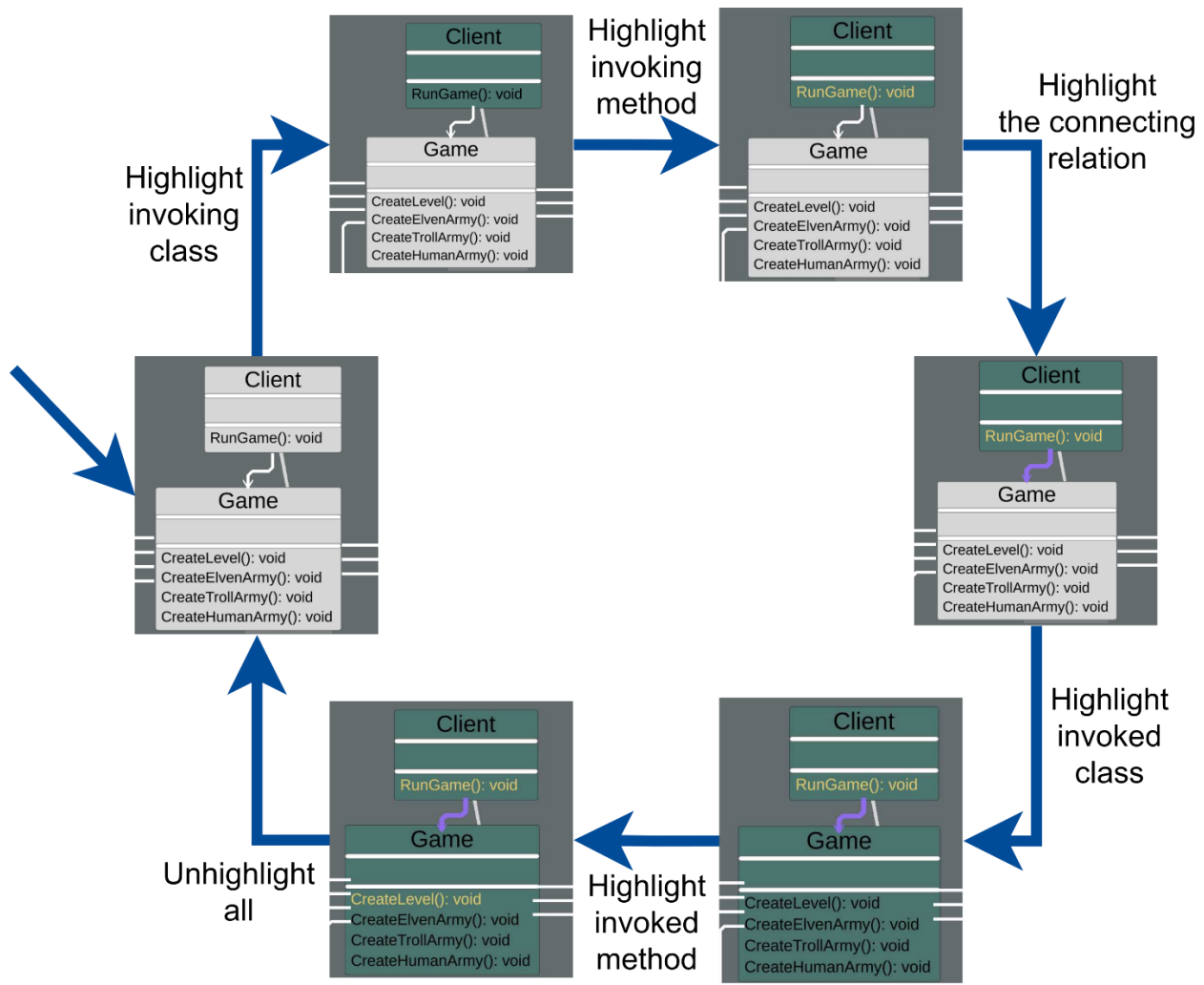

**Figure 3:** Method invocation animation in class diagram layer in our prototype.

### 3.2.1. Class diagram layer

The class diagram layer represents the structure of the modelled object-oriented software system. It utilizes most notation elements of UML class diagram, including classes, relations, attributes and methods.

While method invocation during source code execution is performed by class instances (unless the method is static), animation is projected into class diagram layer. When a class instance invokes a method of another class instance, the corresponding classes are highlighted to indicate where the currently occuring interaction is defined from the structural point of view. Fig. 3 demonstrates how a method invocation is animated in class diagram layer.

### 3.2.2. Object diagram layer

The object diagram layer is not predefined, therefore it is empty at the start of the model execution. During the execution, whenever an object is instantiated, it is reflected into the object diagram layer. As a result, at any given time, the object diagram represents exact state of the runtime, in terms of object instances. All existing object instances, their attribute values and relations are shown in the object diagram layer. The main deviation from UML object diagram stems in listing of methods. To enable animation of method invocation between concrete objects, nodes in object diagram layer also display the methods the object possesses. Each object in the object diagram layer is connected to its corresponding class in the class diagram layer via an inter-diagram connector.

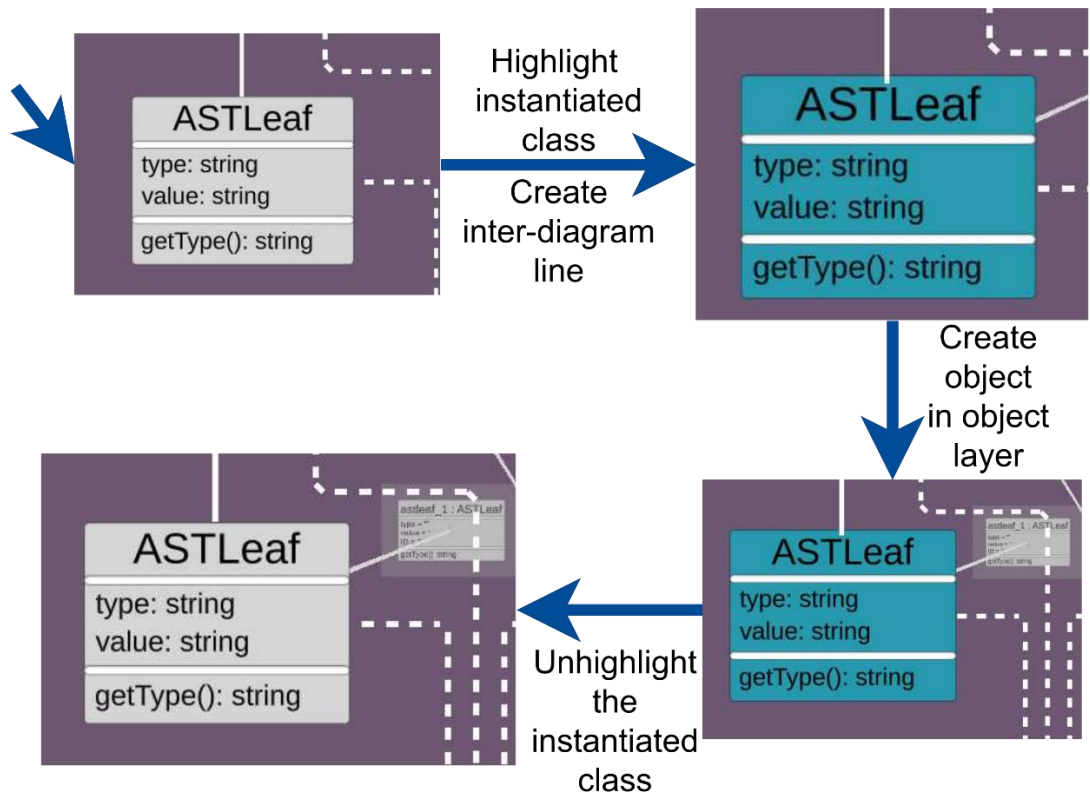

**Figure 4:** Animation of object instantiation in our prototype.

Additionally to the class diagram layer, the object diagram layer is also animated upon a method invocation. Fig. 4 demonstrates how an object instantiation is animated in both class diagram and object diagram layer.

### 3.2.3. Source code layer

The source code layer represents the source code of the currently executed method. The source code of the command that is currently being executed is highlighted. This provides a hint to to the user where to look at in diagrams (e.g. object instantiation command indicates the user to expect a new object in the object diagram layer).  In Fig. 5, we demonstrate invocation of *CreateRanger* method in (a), execution of the method's body in (b), and finally, the *CreateRanger* method execution returns the execution flow in (c).

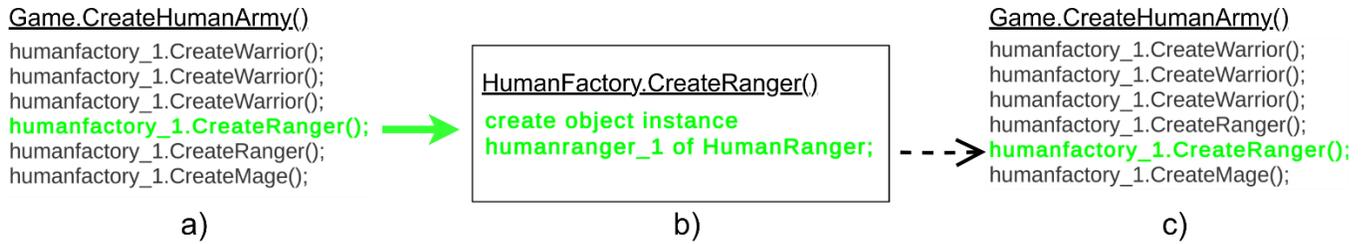

**Figure 5:** Sequential method invocation animation in source code layer in our prototype during model execution.

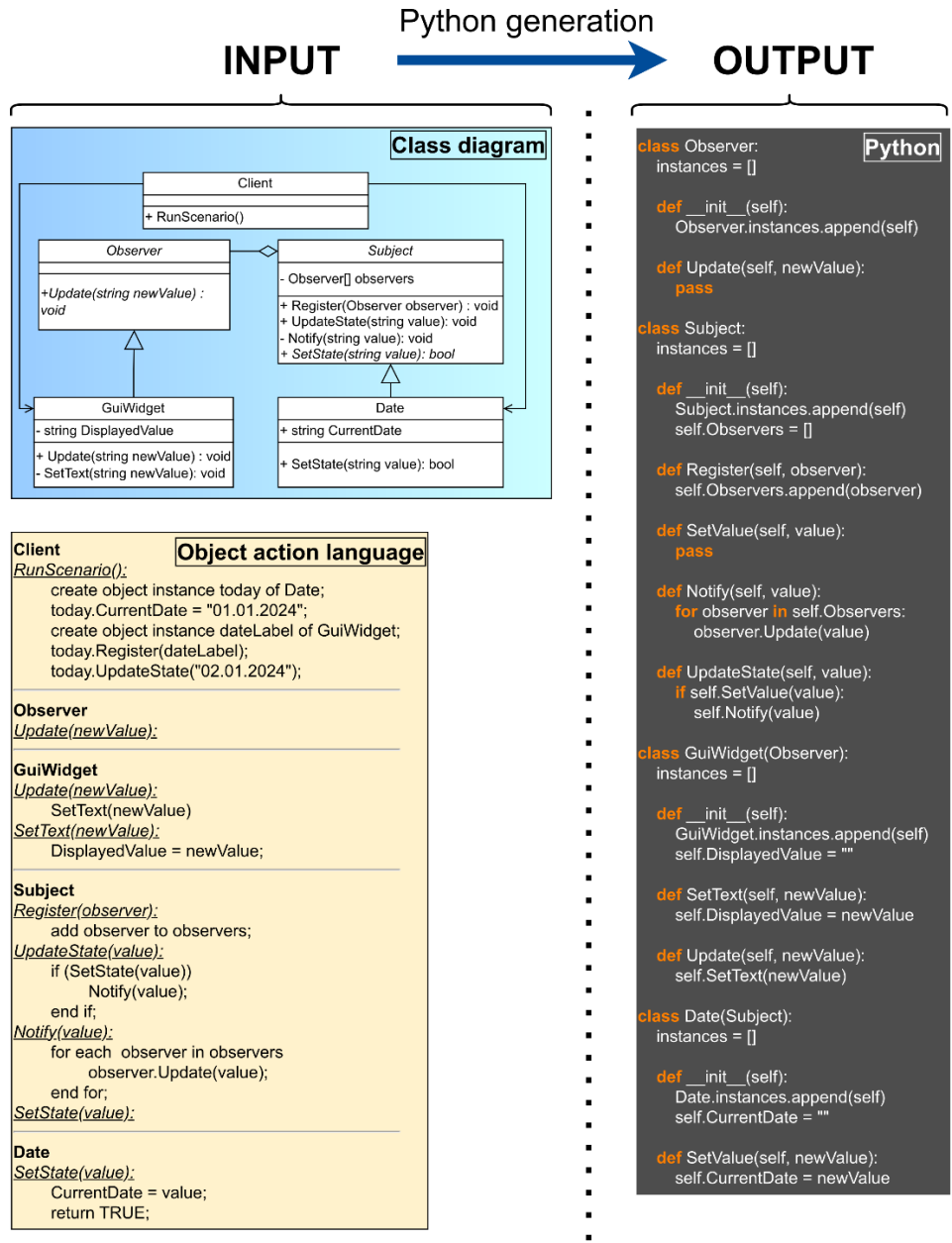

**Figure 6:** Python source code generation example with a very simple Observer pattern implementation.

## 3.3. Generating source code

To demonstrate the intended platform-independent nature of the proposed modelling method, *AnimArch* also supports source code generation. Accepting class diagram and animation source code on input, a platform-dependent source code can be generated. At the present, Python language is the only one supported. The resulting source code is written into a single file, defining all classes and methods from the class diagram. We demonstrate the source code generation process by the class diagram and OAL source code representing input and Python source code representing the corresponding output in Fig. 6.

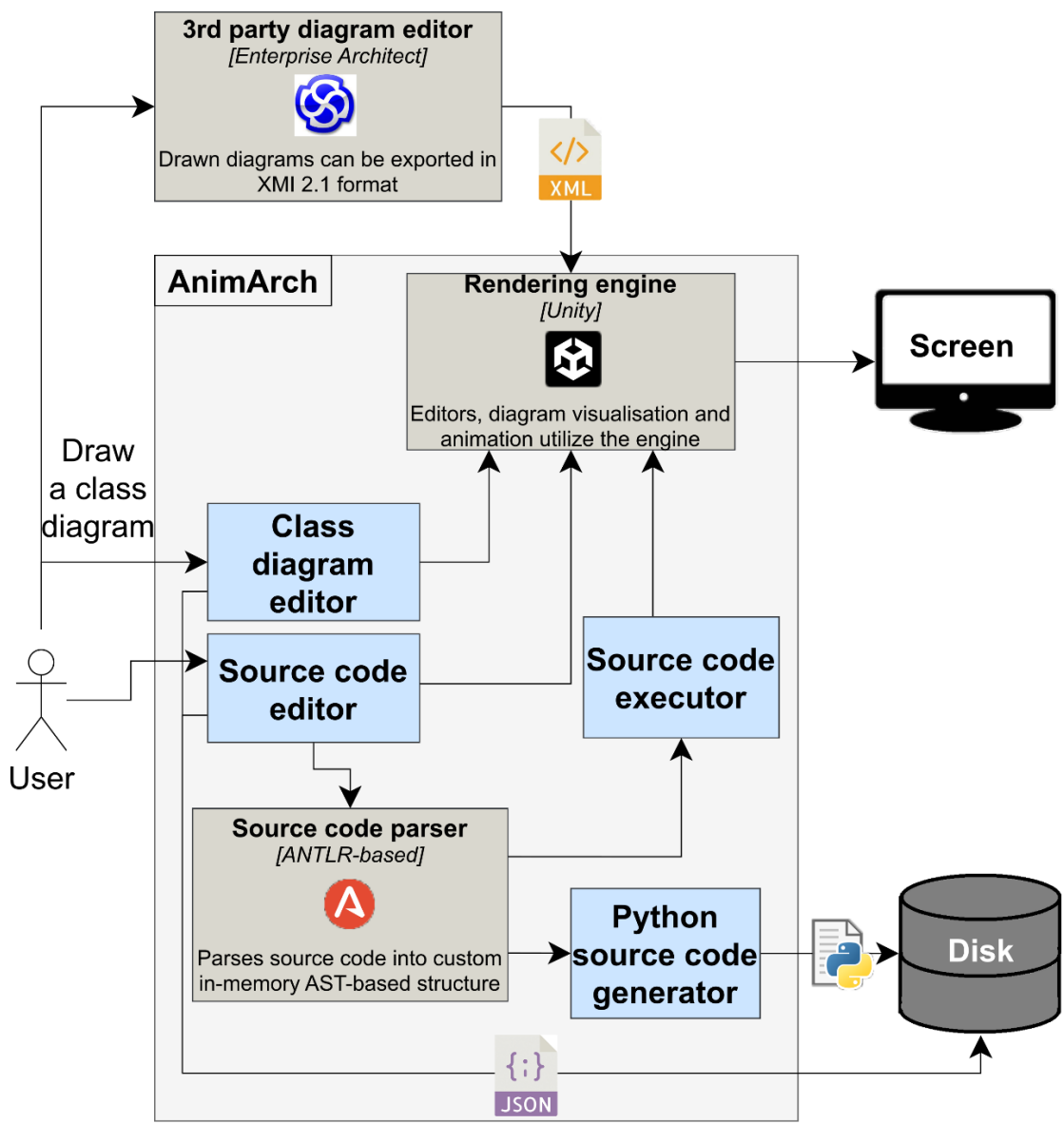

**Figure 7:** Component diagram of our prototype.

## 3.4. Implementation details

*AnimArch* is implemented as a monolithic desktop application utilizing Unity engine[5] for rendering. *AnimArch*'s architecture is shown in Fig. 7. The notation is gently inspired by the C4 model [3] although applied on different abstraction level.

## 4. Conclusion and future work

We have introduced a novel software modelling method and the supporting modelling software, *AnimArch*. Its main properties and features have been demonstrated with several screenshots.

The current objective is to potentially improve user-friendliness of the *AnimArch* software. Afterwards, user evaluation is to be performed to assess viability of *AnimArch* and the proposed modelling method. Currently, visual programming of the OAL source code via sequence diagram drawing is being researched and implemented, as suggested in Fig. 8. Further work might involve improvement of collaboration support, reverse engineering of large software systems, visualization optimizations with regard to scaling, involvement of other UML diagrams, support of network-based collaboration, virtual reality support and AI-based assistance during model creation.

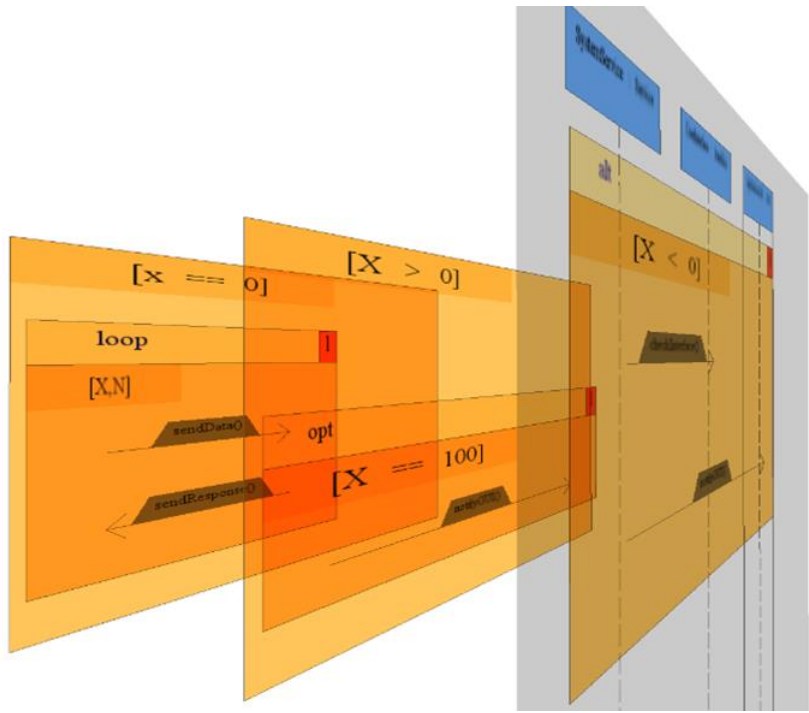

**Figure 8:** Our proposal of visual programming in *AnimArch* using UML sequence diagram.

**ACKNOWLEDGEMENT**. *This work has been carried out in the framework of the TERAIS project, a Horizon-Widera-2021 program of the European Union, GA no. 101079338.*

---

[5] https://unity.com/

# References

[1] Frederick Brooks Jr. "No Silver Bullet Essence and Accidents of Software Engineering". In: IEEE Computer 20 (Apr. 1987), pp. 10–19. doi: 10.1109/MC.1987.1663532.

[2] Keith Brown. "Navigating the rover with xtUML". In: Proceedings of MODELS 2018 Workshops. Vol. 2245. MODELS-WS 2018. Oct. 2018. url: https://ceur-ws.org/Vol-2245/mdetools_paper_10.pdf.

[3] Simon Brown. The C4 model for visualising software architecture. https://leanpub.com/visualising-software-architecture. 2015.

[4] Micaela Esteves and António J Mendes. "OOP-Anim, a system to support learning of basic object-oriented programming concepts." In: CompSysTech. 2003, pp. 573–579.

[5] Matej Ferenc, Ivan Polášek, and Juraj Vincúr. "Collaborative Modeling and Visualization of Software Systems Using Multidimensional UML". In: 2017 IEEE Working Conference on Software Visualization (VISSOFT). 2017, pp. 99–103. doi: 10.1109/VISSOFT.2017.19.

[6] Joseph Gil and Stuart Kent. "Three-Dimensional Software Modelling". In: Proceedings of the 20th International Conference on Software Engineering. ICSE '98. Kyoto, Japan: IEEE Computer Society, 1998, pp. 105–114. isbn: 0818683686.

[7] Lukas Gregorovic and Ivan Polasek. "Analysis and Design of Object-Oriented Software Using Multidimensional UML". In: Proceedings of the 15th International Conference on Knowledge Technologies and Data-Driven Business. i-KNOW '15. Graz, Austria: Association for Computing Machinery, 2015. isbn: 9781450337212. doi: 10.1145/2809563.2809564. url: https://doi.org/10.1145/2809563.2809564.

[8] Steven Halim. "VisuAlgo - Visualising Data Structures and Algorithms Through Animation". In: 2015. url: https://api.semanticscholar.org/CorpusID:65032878.

[9] Devamardeep Hayatpur, Daniel Wigdor, and Haijun Xia. "CrossCode: Multi-level Visualization of Program Execution". In: Proceedings of the 2023 CHI Conference on Human Factors in Computing Systems. CHI '23. ACM, Apr. 2023. doi: 10.1145/3544548.3581390. url: http://dx.doi.org/10.1145/3544548.3581390.

[10] Dean Hendrix et al. "Providing Data Structure Animations in a Lightweight IDE". In: Electronic Notes in Theoretical Computer Science 178 (2007). Proceedings of the Fourth Program Visualization Workshop (PVW 2006), pp. 101–109. issn: 1571-0661. doi: https://doi.org/1 .1016/j.entcs.2007.01.039. url: https://www.sciencedirect.com/science/article/pii/S1571066107002733.

[11] Akihiro Hori, Masumi Kawakami, and Makoto Ichii. "CodeHouse: VR Code Visualization Tool". In: 2019 Working Conference on Software Visualization (VISSOFT). 2019, pp. 83–87. doi: 10.1109/VISSOFT.2019.00018.

[12] Rodi Jolak. "Understanding and Supporting Software Design in Model-Based Software Engineering". In: 2020. url: https://api.semanticscholar.org/CorpusID:213932045.

[13] Rodi Jolak et al. "Software engineering whispers: The effect of textual vs. graphical software design descriptions on software design communication". In: Empirical Software Engineering 25 (2020), pp. 4427–4471. url: https://api.semanticscholar.org/CorpusID:225195437.

[14] Reyhaneh Kalantari and Timothy C. Lethbridge. "Characterizing UX Evaluation in Software Modeling Tools: A Literature Review". In: IEEE Access 10 (2022), pp. 131509–131527. url: https://api.semanticscholar.org/CorpusID:254432745.

[15] Alexander Krause-Glau and Wilhelm Hasselbring. "Collaborative, Code-Proximal Dynamic Software Visualization within Code Editors". In: ArXiv abs/2308.15785 (2023). url: https://api.semanticscholar.org/CorpusID:261339648.

[16] Jakub Kučečka et al. "UML-based Live Programming Environment in Virtual Reality". In: Oct. 2022, pp. 177–181. doi: 10.1109/VISSOFT55257.2022.00028.

[17] N. S. Kumar et al. "Code-Viz: Data Structure Specific Visualization and Animation Tool For User-Provided Code". In: 2021 International Conference on Smart Generation Computing, Communication and Networking (SMART GENCON). 2021, pp. 1–8. doi: 10.1109/SMARTGENCON51891.2021.9645747.

[18] Victor Lian, Elliot Varoy, and Nasser Giacaman. "Learning Object-Oriented Programming Concepts Through Visual Analogies". In: IEEE Transactions on Learning Technologies 15 (2022), pp. 78–92. url: https://api.semanticscholar.org/CorpusID:247179935.

[19] Nathan Mills, Allen Wang, and Nasser Giacaman. "Visual Analogy for Understanding Polymorphism Types". In: Proceedings of the 23rd Australasian Computing Education Conference. ACE '21. Virtual, SA, Australia: Association for Computing Machinery, 2021, pp. 48–57. isbn: 9781450389761. doi: 10.1145/3441636.3442304. url: https://doi.org/10.1145/3441636.3442304.

[20] Roy Oberhauser. "VR-UML: The Unified Modeling Language in Virtual Reality - An Immersive Modeling Experience". In: International Symposium on Business Modeling and Software Design. 2021. url: https://api.semanticscholar.org/CorpusID:237366418.

[21] Johanna Pirker et al. "The Potential of Virtual Reality for Computer Science Education - Engaging Students through Immersive Visualizations". In: 2021 IEEE Conference on Virtual Reality and 3D User Interfaces Abstracts and Workshops (VRW) (2021), pp. 297–302. url: https://api.semanticscholar.org/CorpusID:233990353.

[22] Jorma Sajaniemi, Pauli Byckling, and Petri Gerdt. "Animation Metaphors for Object-Oriented Concepts". In: Electronic Notes in Theoretical Computer Science 178 (2007). Proceedings of the Fourth Program Visualization Workshop (PVW 2006), pp. 15–22. issn: 1571-0661. doi: https://doi.org/10.1016/j.entcs.2007.01.037. url: https://www.sciencedirect.com/science/article/pii/S1571066107002605.

[23] Maxime Savary-Leblanc and Xavier Le Pallec. "Interactive Highlighting for Digital UML Class Diagrams: A New Feature". In: Proceedings of the 25th International Conference on Model Driven Engineering Languages and Systems: Companion Proceedings. MODELS '22. Montreal, Quebec, Canada: Association for Computing Machinery, 2022, pp. 247–256. isbn: 9781450394673. doi: 10.1145/3550356.3561557. url: https://doi.org/10.1145/3550356.3561557.

[24] P. Sparsha, Pawan Harish, and N. S. Kumar. "Visualization of Data Structures with Animation of Code". In: Innovative Data Communication Technologies and Application (2021). url: https://api.semanticscholar.org/CorpusID:234335100.

[25] Martin Stancek et al. "Collaborative software design and modeling in virtual reality". In: Information and Software Technology 166 (2024), p. 107369. issn: 0950-5849. doi: https://doi.org/10.1016/j.infsof.2023.107369. url: https://www.sciencedirect.com/science/article/pii/S0950584923002240.

[26] Boban Vesin, Rodi Jolak, and Michel R.V. Chaudron. "OctoUML: An Environment for Exploratory and Collaborative Software Design". In: 2017 IEEE/ACM 39th International Conference on Software Engineering Companion (ICSE-C). 2017, pp. 7–10. doi: 10.1109/ICSE-C.2017.19.

[27] Jeong Yang, Young Lee, and Kai-Hsiung Chang. "Evaluations of JaguarCode: A web-based object-oriented programming environment with static and dynamic visualization". In: J. Syst. Softw. 145 (2018), pp. 147–163. url: https://api.semanticscholar.org/CorpusID:52965006.

[28] Enes Yigitbas et al. "Collaborative Software Modeling in Virtual Reality". In: CoRR abs/2107.12772 (2021). arXiv: 210.12772. url: https://arxiv.org/abs/2107.12772.

[29] Enes Yigitbas et al. "Design and evaluation of a collaborative UML modeling environment in virtual reality". In: Software and Systems Modeling (2022), pp. 1–29.